\documentclass[12pt]{article}
\usepackage{fullpage,citesort,
epsfig,psfrag,graphics,amsbsy,amssymb,amsmath,
}
\usepackage{caption}
\newcommand{\beq}{\begin{equation}}
\newcommand{\eeq}{\end{equation}}
\newcommand{\bea}{\begin{eqnarray}}
\newcommand{\eea}{\end{eqnarray}}
\newcommand{\bfs}{\boldsymbol}

\newcommand{\Tr}{{\rm Tr}}
\newcommand{\tr}{{\rm tr}}
\newcommand{\ket}[1]{|#1\rangle}
\newcommand{\bra}[1]{\langle#1|}

\def\math{\mathsurround=0pt }
\def\leftrightarrowfill{$\math \mathord\leftarrow \mkern-6mu
 \cleaders\hbox{$\mkern-2mu \mathord- \mkern-2mu$}\hfill
 \mkern-6mu \mathord\rightarrow$}
\def\overleftrightarrow#1{\vbox{\ialign{##\crcr
     \leftrightarrowfill\crcr\noalign{\kern-1pt\nointerlineskip}
     $\hfil\displaystyle{#1}\hfil$\crcr}}}

\let\l=\lambda

 \def\bd{\begin{document}} \def\ed{\end{document}}
\def\ds{\documentstyle} \let\fr=\frac \let\bl=\bigl \let\br=\bigr
\let\Br=\Bigr \let\Bl=\Bigl
\let\bm=\bibitem
\let\na=\nabla
\let\pa=\partial \let\ov=\overline
\def\ft#1#2{{\textstyle{{\scriptstyle #1}\over {\scriptstyle #2}}}}
\def\fft#1#2{{#1 \over #2}}
\def\vp{\varphi}
\def\sst#1{{\scriptscriptstyle #1}}
\def\oneone{\rlap 1\mkern4mu{\rm l}}
\def\td{\tilde}
\def\wtd{\widetilde}
\def\dalemb#1#2{{\vbox{\hrule height .#2pt
        \hbox{\vrule width.#2pt height#1pt \kern#1pt
                \vrule width.#2pt}
        \hrule height.#2pt}}}
\def\square{\mathord{\dalemb{6.8}{7}\hbox{\hskip1pt}}}
\def\wtd{\widetilde}
\def\R{\rlap{\rm I}\mkern3mu{\rm R}}
\def\im{{\rm i}}
\def\tilg{\tilde{g}}
\def\tilF{\tilde{F}}
\def\tilA{\tilde{A}}
\def\varf{\varphi}
\def\tilf{\tilde{\phi}}
\def\tilh{\tilde{h}}
\def\rme{{\rm e}}
\def\ep{\epsilon}
\def\0{{(0)}}
\def\9{{(9)}}
\def\8{{(8)}}
\def\7{{(7)}}
\def\6{{(6)}}
\def\5{{(5)}}
\def\4{{(4)}}
\def\3{{(3)}}
\def\2{{(2)}}
\def\1{{(1)}}
\newcommand{\trace}{{\rm Tr}}
\newcommand{\ub}{\overline{U}}
\newcommand{\vb}{\overline{V}}
\newcommand{\uh}{\widehat{U}}
\newcommand{\vh}{\widehat{V}}
\newcommand{\ubh}{\overline{\widehat{U}}}
\newcommand{\vbh}{\overline{\widehat{V}}}
\newcommand{\lb}{\bar{\l}}
\newcommand{\Fb}{\overline{F}}
\newcommand{\Fh}{\widehat{F}}
\newcommand{\Fbh}{\overline{\widehat{F}}}
\newcommand{\Ab}{\overline{A}}
\newcommand{\Ah}{\widehat{A}}
\newcommand{\Abh}{\overline{\widehat{A}}}
\newcommand{\Gb}{\overline{G}}
\newcommand{\Gh}{\widehat{G}}
\newcommand{\Gbh}{\overline{\widehat{G}}}
\newcommand{\Pb}{\overline{P}}
\newcommand{\Ph}{\widehat{P}}
\newcommand{\Pbh}{\overline{\widehat{P}}}
\newcommand{\Qb}{\overline{Q}}
\newcommand{\Qh}{\widehat{Q}}
\newcommand{\Qbh}{\overline{\widehat{Q}}}
\newcommand{\Bb}{\overline{B}}
\newcommand{\Bh}{\widehat{B}}
\newcommand{\Bbh}{\overline{\widehat{B}}}
\newcommand{\fhns}{\hat{F}^{\rm (NS)}}
\newcommand{\fhrr}{\hat{F}^{\rm (RR)}}
\newcommand{\ahns}{\hat{A}^{\rm (NS)}}
\newcommand{\ahrr}{\hat{A}^{\rm (RR)}}
\newcommand{\hhrr}{\hat{H}^{\rm (RR)}}
\newcommand{\hchi}{\hat{\chi}}
\newcommand{\hphi}{\hat{\phi}}
\newcommand{\htau}{\hat{\tau}}
\newcommand{\cG}{{\cal G}}
\newcommand{\cGb}{\overline{{\cal G}}}
\newcommand{\cH}{{\cal H}}
\newcommand{\cP}{{\cal P}}
\newcommand{\cPb}{\overline{{\cal P}}}
\newcommand{\cQ}{{\cal Q}}
\newcommand{\cQb}{\overline{{\cal Q}}}
\newcommand{\cM}{{\cal M}}
\newcommand{\cN}{{\cal N}}
\newcommand{\cO}{{\cal O}}
\newcommand{\cD}{{\cal D}}
\newcommand{\cL}{{\cal L}}

\newcommand{\vpp}{\mbox{$\langle{\scriptstyle++}\rangle$}}
\newcommand{\vmp}{\mbox{$\langle{\scriptstyle-+}\rangle$}}
\newcommand{\vppp}{\mbox{$\langle{\scriptstyle+++}\rangle$}}
\newcommand{\vmpp}{\mbox{$\langle{\scriptstyle-++}\rangle$}}
\newcommand{\vpmp}{\mbox{$\langle{\scriptstyle+-+}\rangle$}}

\begin{document}
\setlength{\captionmargin}{36pt}
\setlength{\captionmargin}{36pt}
\begin{titlepage}
\begin{flushright}
\phantom{UFIFT-HEP}
\end{flushright}

\vskip 3cm
\begin{center}
\begin{large}
{\bf $1/N$ Perturbations in Superstring Bit Models}
\end{large}

\vskip 2cm
{\large
 Charles B. Thorn\footnote{E-mail  address: {\tt thorn@phys.ufl.edu}}
}
\vskip0.20cm
{\it Institute for Fundamental Theory,\\
Department of Physics, University of Florida,
Gainesville FL 32611}


\vskip 1.0cm
\end{center}

\begin{abstract}
\noindent We develop the $1/N$ expansion for stable string bit models,
focussing on a model with bit creation operators 
carrying only transverse spinor indices $a=1,\ldots,s$. At leading order
($N=\infty$), this model produces a (discretized) 
lightcone string with a ``transverse
space'' of $s$ Grassmann worldsheet fields. 
Higher orders in the $1/N$ expansion are shown to be
determined by the overlap of a single large closed chain (discretized string)
with two smaller closed chains. In the models studied here, the overlap
is not accompanied with operator insertions at the break/join point.
Then the requirement that the discretized overlap have a smooth continuum
limit leads to the critical Grassmann ``dimension'' of $s=24$. This 
``protostring'', a Grassmann analog of the bosonic string, 
is unusual, because it has no large transverse dimensions. 
It is a string moving in one space dimension
and there are neither tachyons nor massless particles. 
The protostring, derived from our pure spinor string bit model,
has 24 Grassmann dimensions, 16 of which could be bosonized to form
8 compactified bosonic dimensions, leaving 8 Grassmann dimensions--
the worldsheet content of the superstring. If the transverse space 
of the protostring could be ``decompactified'',
string bit models might provide an appealing and solid foundation for 
superstring theory.
\end{abstract}
\vfill
\end{titlepage}
\section{Introduction}
When string theory is formulated in lightcone coordinates
($x^\pm=(x^0\pm x^1)/\sqrt{2}$) \cite{goddardgrt}, it is
natural to propose that string is a composite of elementary entities
called string bits \cite{gilest,thornsakh}. Incorporating
supersymmetry in string bit models, leads to the
proposal \cite{bergmantsubit} that the superstring bit creation operator 
has the structure
\bea
({\bar\phi}_{[a_1\cdots a_n]})_\alpha^{\ \beta}({\bfs x}),\qquad a_i=1,\ldots,s,
\qquad n=0,\ldots, s,\qquad \alpha, \beta=1,\ldots,N,
\eea
where ${\bfs x}$ denotes the transverse coordinates of the lightcone,
and the square brackets in the subscript remind us that the $a_i$'s
are completely antisymmetric. The $a_i$'s are spinor indices of the
transverse rotation group, and $\alpha,\beta$ are color indices, which
are introduced to formulate a dynamics that favors the formation of
long (closed) chains of string bits. The bit number operator 
$M=\sum_n\tr({\bar \phi}_{[a_1\cdots a_n]}\phi_{[a_1\cdots a_n]}/n!)$ is naturally 
identified with the `$+$' component
of momentum $P^+=(P^0+P^1)/\sqrt{2}=mM$. Chains with
$M\to\infty$ would then become continuous closed strings. It is
central to the string bit hypothesis that string bits are not 
{\it a priori} confined in chains but that chain formation is a consequence
of the dynamics. Such  dynamics will arise generically in the 't Hooft
$N\to\infty$ limit \cite{thooftlargen}. 
In this original formulation of string bit models
there is an aspect of 't Hooft's idea of holography 
\cite{thoofthologram} in that the
the fundamental string bits only move in the transverse space, while the
strings behave as though moving in transverse space plus an extra 
spatial dimension $x^-$, the canonical conjugate of $P^+$.

Recently, we have noted that the transverse coordinates are
extraneous, and this led to the proposal 
that the bits have no space to move at all \cite{sunthorn}. 
This proposal is a rather more drastic form of holography 
in which all space dimensions,
and not just the longitudinal one, emerge from the dynamics of string
formation from string bits \cite{thornspace}. The idea is that, with 
suitable dynamics, some spin degrees of freedom carried by the string bit
can, on long chains, fluctuate collectively as one dimensional spin waves. 
Such spin waves are well known to act as a compactified bosonic coordinate.
In these two papers the string bit creation operator is then taken to
be the simpler
\bea
({\bar\phi}_{a_1\cdots a_n})_\alpha^{\ \beta}&&,\qquad a_i=1,\ldots,s,
\qquad n=0,\ldots, s,\qquad \alpha, \beta=1,\ldots,N,
\label{bitphis}
\eea
where here and from now on, we suppress the square brackets around
the spinor indices. 
In \cite{thornspace} a further set of ``flavor'' indices is appended to the
$\phi$'s to serve as the seed for the transverse coordinates. However, we
refrain from adding them here, because we hope that the seeds for 
transverse space can somehow be found by enlarging the value of $s$.
Fluctuations in the $a$'s produce, on long chains, 
the Grassmann worldsheet fields
$\theta^a_{L,R}$ of the Green-Schwarz type \cite{greenschwarz}. 
If $s=24$, eight of the $\theta$'s
could take the role of the superstring Grassmann fields, but the remaining
16 could be bosonized into 8 (compactified) transverse coordinates.

In this article our aim is to study, by perturbing in $1/N$, not only
the precise manner
in which string bit dynamics lead to the free superstring 
spectrum at $N=\infty$, but also how the $1/N$ corrections lead to the
three string interaction vertex of string theory. We will start in Section 2
by setting up the systematic $1/N$ perturbation expansion in string bit models.
We will see that the structure of this perturbation theory follows that
of Mandelstam's interacting string diagrams \cite{mandelstamlc}. Then we
proceed to apply this formalism to our pure spinor string bit model,
which we also call the protostring bit model.
In Section 3 we obtain the exact zeroth order spectrum (at $N=\infty$).
In Section 4 we discuss the calculation of the overlap that describes the
cubic vertex. We formulate the overlap calculation for finite bit number chains,
and then discuss the continuum limit in which the bit numbers of each
string tend to infinity at fixed ratio. These results are aided by 
numerical calculations using MATLAB. We compare our conclusions
with those in the literature for various situations. In Appendix C,
we present an analytic computation of the overlap in the continuum limit.
Then we use this result to determine how various operator insertions
behave when inserted at the break/join point of the string
to 2 string transition. Finally, in Section 5 we put everything together
to construct the total vertex, respecting the requirement that
the physical amplitudes have a finite continuum limit. As seen
in \cite{gilest}, this last requirement determines the critical dimension.
We list and compare the ways in which this requirement is met for
 the bosonic string, for the
IIB superstring \cite{gso,greenschwarz}, for the RNS string \cite{rns}, 
and finally for the new
``protostring'' which is the outcome of our pure spinor superstring bit
model. We close with a discussion of the properties of the
protostring, which has 24 Grassmann dimensions, has no transverse
bosonic dimensions, and is expected
to have a spectrum with no massless particles, i.e. that it possesses a
mass gap. We include three appendices containing technical details
to supplement the main text.

\section{$1/N$ Expansion}
The string bit model we focus on in this article takes as fundamental
variables the creation operators of (\ref{bitphis}). Their
(anti)commutation relations are given in Appendix A.
We shall keep
$s$ a general positive integer, and we shall analyze the Hamiltonian $H_S$
given in \cite{thornspace} and quoted in detail
(see Eq.(\ref{sham})) in
Appendix A, along with its action on color singlet states. To guide
the reader's eye we display here the $s=1$ Hamiltonian: 
\bea
H^{s=1}&=&\frac{2}{N}\Tr\left[({\bar a}^2 -i{\bar b}^2)a^2
-({\bar b}^2-i{\bar a}^2)b^2+({\bar a} {\bar b}
+{\bar b} {\bar a})ba+({\bar a} {\bar b}
-{\bar b} {\bar a})ab\right].
\label{bitham}
\eea
In this special case there is one bosonic bit $\phi=a$ and one fermionic
bit $\phi_1=b$. The Hamiltonian for general $s$ is a good deal more complex.

For the analysis to follow, it will be convenient to introduce Grassmann
coordinates $\theta^a$, $a=1,\ldots,s$ and define a super bit
creation operator
\bea
\psi(\theta)&=&\sum_{k=0}^s
\frac{1}{k!}{\bar\phi}_{c_1\cdots c_k}\theta^{c_1}\cdots\theta^{c_k}.
\eea
The ${\bar\phi}$ can be recovered from $\psi$ recursively by taking multiple
Grassmann derivatives, starting with $s$ derivatives which singles out
${\bar\phi}_{a_1\cdots a_s}$, Then single out $\phi_{a_1\cdots a_{s-1}}$ by 
applying $s-1$ derivatives on $\psi$ minus the
contribution of ${\bar\phi}_{a_1\cdots a_s}$, and so on. 
To work on the color singlet subspace
of Fock space, we define an empty state $\ket{0}$ and the set of trace operators
\bea
T(\theta_1,\ldots,\theta_k)=\Tr\psi(\theta_1)\cdots\psi(\theta_k),
\eea
where the $\theta$'s are  $s$-component Grassmann variables.
Then the color singlet subspace is spanned by states of the form
\bea
T(\theta_1,\ldots,\theta_K)T(\eta_{1},\ldots,\eta_{L})\cdots\ket{0}.
\eea
\subsection{Action of $H$ on Multi-Trace States}
In Appendix A we present the Hamiltonian as the sum of five
terms and give the action of each term on color singlet states..
We can summarize the action of $H=\sum_{i=1}^5H_i$
on multitrace states by defining
\bea
{\bar h}_{kl}&=&2\left(s-2\theta_k^a\frac{d}{d\theta_k^a}\right)
+2\theta_k^a\frac{d}{d\theta_{l}^a}+2\theta_l^a\frac{d}{d\theta_{k}^a}
-2i\theta_k^a{\theta_{l}^a}-2i\frac{d}{d\theta_k^a}\frac{d}{d\theta_{l}^a}\\
{\bar h}&=&\sum_{k=1}^M{\bar h}_{k,k+1}.\eea
Then, there follows
\bea
HT(\theta_1\cdots\theta_M)\ket{0}&=&{\bar h}T(\theta_1\cdots\theta_M)\ket{0}
\nonumber\\
&&+\frac{1}{N}\sum_{k,l\neq k,k+1}{\bar h}_{kl}T(\theta_l\cdots\theta_k)
T(\theta_{k+1}\cdots\theta_{l-1})\ket{0}\\
HT({\theta_1\cdots\theta_K})
T({\eta_{1}\cdots\eta_L})\ket{0}&=&({\bar h}_\theta+{\bar h}_\eta)
T({\theta_1\cdots\theta_K})
T({\eta_{1}\cdots\eta_L})+\frac{1}{N}{\rm Fission~Terms}\nonumber\\
&& +\frac{1}{N}\sum_{k=1}^K\sum_{l=1}^L{\bar h}_{kl}
T(\theta_{k+1}\cdots\theta_{k}\eta_l\cdots\eta_{l-1})\ket{0}
\nonumber\\
&&+\frac{1}{N}\sum_{k=1}^K\sum_{l=1}^L  
{\bar h}_{lk}T(\theta_k\cdots\theta_{k-1}{\eta_{l+1}
\cdots\eta_{l}})\ket{0}.
\eea
In the development of perturbation theory,
we shall transfer the derivatives of ${\bar h}_{kl}$ to act on the 
coefficient amplitude
multiplying each multitrace state, whence
it will take the form
\bea
{h}_{kl}&=&-2\left(s-2\theta_k^a\frac{d}{d\theta_k^a}\right)
-2\theta_k^a\frac{d}{d\theta_{l}^a}-2\theta_l^a\frac{d}{d\theta_{k}^a}
-2i\theta_k^a{\theta_{l}^a}-2i\frac{d}{d\theta_k^a}\frac{d}{d\theta_{l}^a}\\
{h}&=&\sum_{k=1}^M{h}_{k,k+1}.\eea
We shall also make use of Grassmann variables that satisfy a Clifford
algebra:
\bea
S^a_k&=&\theta_k^a+\frac{d}{\theta_k^a},\qquad {\tilde S}_k^a=
i\left(\theta_k^a-\frac{d}{\theta_k^a}\right)\\
\{S_k^a,S_l^b\}&=&2\delta_{kl}\delta_{ab},\qquad 
\{{\tilde S}_k^a,{\tilde S}_l^b\}=2\delta_{kl}\delta_{ab},\qquad
\{S_k^a,{\tilde S}_l^b\}=0.
\eea
Then $h_{kl}$ becomes
\bea
h_{kl}&=&-iS^a_kS^a_l+i{\tilde S}^a_k{\tilde S}^a_l-iS_k^a{\tilde S}^a_l
+i{\tilde S^a}_kS^a_l+2iS^a_k{\tilde S}^a_k.
\eea 
\subsection{Systematic Perturbation theory}
We develop the $1/N$ expansion on Fock space, following the methods of
\cite{thornfock}. 
At zeroth order the first task is to solve the eigenvalue problem
\bea
h\psi_r(\theta_1,\cdots,\theta_{M_r})&=& 
E_r\psi_r(\theta_1,\cdots,\theta_{M_r}),
\eea
and then we change the single trace operators to energy basis
\bea
T_r&=&\int d^s\theta_1\cdots d^s\theta_{M_r} T(\theta_1,\cdots,\theta_{M_r})
\psi_r(\theta_1,\cdots,\theta_{M_r}).
\eea
Because the $T$ are cyclically symmetric we may assume that the $\psi_r$
satisfy the cyclic property
\bea
\psi_r(\theta_1,\cdots,\theta_{M_r})=(-)^{s(M-1)}
\psi_r(\theta_2,\cdots,\theta_{M_r},\theta_1).
\eea
The potential minus sign is due to the fact that if $s$ and $M-1$ are odd, the
cyclic transform of the measure acquires a minus sign.

Define the conjugate to $\psi_r$, denoted ${\bar\psi}_r$, such that
\bea
\int d\theta_1\cdots d\theta_{M_r} {\bar\psi}_s(\theta_1,\cdots,\theta_{M_r})
\psi_r(\theta_1,\cdots,\theta_{M_r})=\delta_{rs},
\eea
and so the completeness relation is written
\bea
\sum_r \psi_r(\theta_1,\cdots,\theta_{M_r}){\bar\psi}_r(\phi_1,\cdots,\phi_{M_r})
=\delta(\theta-\phi),
\eea
where the delta function is understood to be symmetrized under
cyclic permutations. In the energy basis the action of $H$ 
on a single trace state becomes
\bea
HT_r\ket{0}&=&E_rT_r\ket{0}+\frac{1}{N}\int d\theta
\sum_{l\neq k,k+1}{\bar h}_{kl}T({\theta_l\cdots\theta_{k}})T({\theta_{k+1}
\cdots\theta_{l-1}})\ket{0}\psi_r(\theta_1,\cdots,\theta_M)\nonumber\\
&\equiv&E_rT_r\ket{0}+\frac{1}{N}\sum_{s,t}T_s T_t\ket{0}V_{str}\nonumber\\
V_{str}&=&\sum_{l\neq k,k+1}\int d\theta {\bar\psi_s}({\theta_l\cdots\theta_{k}}) {\bar\psi_t}({\theta_{k+1}\cdots\theta_{l-1}}){h}_{kl}\psi_r\theta_1,\cdots,\theta_M).
\eea
We see that $h_{kl}$ acts to the left on the eigenfunction ${\bar\psi}_s$,
in which $k,l$ label nearest neighbor pairs of $\theta$'s. We
can normal order  $h_{kl}$ and get the normal ordering constant by
calculating
\bea
\alpha_{kl}&=&\bra{\psi_G}h_{kl}\ket{\psi_G}=
\frac{1}{M_s}\bra{\psi_G}h\ket{\psi_G}=\frac{E_G}{M_s},
\eea
where the second equality follows because ${\bar\psi_G}$ is cyclically
invariant. Thus each term of $h=\sum_kh_{k,k+1}$ contributes an equal amount.
In the continuum limit $E_G\sim \alpha M_s+O(1/M_s)$, so that in this
limit $\alpha_{kl}=\alpha$. Thus we have
\bea
\bra{\psi_s}h_{kl}&=&\bra{\psi_s}(:h_{kl}:+\alpha).
\eea
The terms in the operator $:h_{kl}:$ are nominally of order $M_s^{-1}$,
and so they nominally vanish in the continuum limit. However, it can
be shown (see Appendix C) that the energy lowering components of 
$S_k$, $S_l$, ${\tilde S}_k$,
or ${\tilde S}_l$, nominally of order
$M_s^{-1/2}$,  acting to the right
give a Grassmann odd factor $S$  of order $1$ in the continuum limit
(independently of which operator is chosen). 
In other words, the singularity at the joining point can produce a
factor $M_s^{1/2}$ for each $S_k$. Thus the terms in $:h_{kl}:$ with
two such lowering operators can potentially contribute at order $1$. Happily
the contribution is $S^2=0$ because $S$ is Grassmann odd! 
Thus in the continuous string limit of our
model, the vertex is a pure overlap with no
operator insertions at the joining point: 
\bea
V_{str}&\sim&\alpha\sum_{l\neq k,k+1}\int d\theta 
{\bar\psi_s}({\theta_l\cdots\theta_{k}}) 
{\bar\psi_t}({\theta_{k+1}\cdots\theta_{l-1}})\psi_r(\theta_1,\cdots,\theta_M).
\eea
The fission operation
on any multi-trace state acts on each trace factor just as shown above..
On multi-trace states, the Hamiltonian can also fuse any pair of 
traces into one as follows
\bea
HT_sT_t\ket{0}&\equiv&(E_s+E_t)T_sT_t\ket{0}+\frac{1}{N}T_r\ket{0}W_{rst}
+\frac{1}{N}(T_uT_vT_tV_{uvs}+T_sT_uT_vV_{uvt})\ket{0}.
\eea
The second term on the right is the fusion term arising from
\bea
T_r\ket{0}W_{rst}&=&\int d\theta d\phi
\sum_{k=1}^{M_s}\sum_{l=1}^{M_t}\left[{\bar h}_{kl}T(\theta_{k+1}\cdots\theta_k
\phi_l\cdots\phi_{l-1})\right.\nonumber\\
&&\left.+{\bar h}_{lk}T(\phi_{l+1}\cdots\phi_l
\theta_k\cdots\theta_{k-1})\right]\psi_s(\theta_1\cdots\theta_{m_s})
\psi_t(\phi_1\cdots\phi_{M_t}),\eea
from which we infer
\bea
W_{rst}&=&\sum_{k=1}^{M_s}\sum_{l=1}^{M_t}
\int d\theta d\phi\bigg[{\bar\psi}_r(\theta_{k+1}\cdots\theta_k
\phi_l\cdots\phi_{l-1})h_{kl}\psi_s(\theta_1\cdots\theta_{m_s})
\psi_t(\phi_1\cdots\phi_{M_t})\nonumber\\
&&+{\bar\psi}_r(\phi_{l+1}\cdots\phi_l
\theta_k\cdots\theta_{k-1})h_{lk}\psi_s(\theta_1\cdots\theta_{m_s})
\psi_t(\phi_1\cdots\phi_{M_t})\bigg].
\eea
Again, in the continuum limit $h_{kl}$ and $h_{lk}$ can be replaced
by $\alpha$, in which case the two terms are equal, giving a net
factor of two.

The double sums in $V$ and $W$ have the simple interpretation of
including all ways of splitting a chain in two or of joining
two chains into one. In the first case one picks two bits where the
split takes case. In the second case one must pick a bit on each chain
where the two chains join. Since these events can happen with any pair of
bits, one must sum over all choices. In the continuum limit, these
double sums should go over to double integrals $\sum_{k,l}\to
(1/m^2)\int d\sigma d\sigma^\prime$, and since the factor $\alpha$ 
includes a factor $(1/m)$, the overlap should supply a
factor of $1/M^3$ to get a finite continuum limit. 

As an application, consider the energy eigenvalue problem in perturbation 
theory. Start by expanding the sought eigenstate in trace states:
\bea
\ket{E}&=& \sum_r T_r\ket{0}C^1_r+ \sum_{st} T_sT_t\ket{0}C^2_{st}
+\sum_{stu} T_sT_tT_u\ket{0}C^3_{stu}+\cdots,\eea
and require that $(H-E)\ket{E}=0$:
\bea
0&=&\sum_r (E_r-E)T_r\ket{0}C^1_r+ \sum_{st}(E_s+E_t -E)T_sT_t\ket{0}C^2_{st}
\nonumber\\
&&+\sum_{stu}(E_s+E_t+E_u-E)T_sT_tT_u\ket{0}C^3_{stu}\nonumber\\
&&+\frac{1}{N}\sum_{str}T_sT_t\ket_{0}
V_{str}C^1_r+\frac{1}{N}\sum_{rst}T_r\ket{0}W_{rst}C^2_{st}+\cdots .
\eea
Then equating coefficients of like terms, we have the sequence of equations
\bea
(E_r-E)C^1_r+\frac{1}{N}\sum_{st}W_{rst}C^2_{st}&=&0\\
(E_s+E_t -E)C^2_{st}+\frac{1}{N}\sum_{r}V_{str}C^1_r
+C^3~{\rm Terms}&=&0,
\eea
and so on. For example we can choose the $C^1_r$ with common $E_r$ to be
nonzero at zeroth order and all other $C$'s zero at aeroth order.. 
Then the $C_3$ terms in the
second equation are of order $1/N^2$, so we obtain
\bea
C^2_{st}&=&\frac{1}{E-E_s-E_t}\frac{1}{N}\sum_{r}V_{str}C^1_r
+O(N^{-2})\\
(E-E_r)C^1_r&=&
\frac{1}{N^2}\sum_{st}W_{rst}\frac{1}{E-E_s-E_t}
\sum_{u}V_{stu}C^1_u+O(N^{-3}).\eea
In the $M\to\infty$ limit, when the
energy eigenvalues become continuous, the first equation 
can be interpreted as the amplitude for a single
string to decay into two strings . For any finite $M$, the second 
equation shows that the eigenvalues of the matrix
\bea
\Delta_{ru}&=&\frac{1}{N^2}\sum_{st}W_{rst}\frac{1}{E_r-E_s-E_t}V_{stu}
\eea
determine the lowest order energy shifts to the level $E_r$.

\section{Diagonalizing $h$}
In the preceding section, we have shown how the $1/N$ expansion of
string bit models is determined by what we might call first quantized
string calculations. The legacy of the underlying string bit models
for these calculations is essentially the provision of a fundamental
cutoff, namely the interpretation of a continuous $P^+$ by the 
discrete bit number. Finding the
eigenvalues of $h$ is straightforward, because $h$ is bilinear in
Clifford variables. Therefore, it can be solved by finding energy raising and
lowering operators. This was done in \cite{bergmantsubit}, so we just
give here a quick sketch of the procedure and results. Because $h$ is
the sum of terms $h_a$ each of which contain only the variables with
component $a$, it suffices to work with just one component. In the following,
we suppress the spinor index.

To begin, it is convenient to introduce Fourier transforms:
\bea
B_n&=&\frac{1}{\sqrt{M}}\sum_{k=1}^MS_k e^{-2\pi ikn/M},\qquad
{\tilde B}_n=\frac{1}{\sqrt{M}}\sum_{k=1}^M{\tilde S}_k e^{-2\pi ikn/M}\\
S_k&=&\frac{1}{\sqrt{M}}\sum_{n=0}^{M-1}B_n e^{2\pi ikn/M},\qquad
{\tilde S}_k=\frac{1}{\sqrt{M}}\sum_{n=0}^{M-1}{\tilde B}_n e^{2\pi ikn/M}\\
\{B_m,B_n\}&=&2\delta_{m+n,M},\qquad \{{\tilde B}_m,{\tilde B}_n\}=2\delta_{m+n,M},
\qquad \{{\tilde B}_m,{B}_n\}=0.
\eea
Then we can express $h$ in terms of these
\bea
h&=&\sum_{n=1}^{M-1}\left[-iB_{M-n}B_n e^{2\pi in/M}
+i{\tilde B}_{M-n}{\tilde B}_ne^{2\pi in/M}
+2iB_{M-n}{\tilde B}_n\left(1-\cos\frac{2\pi n}{M}\right)\right].
\eea
We search for eigenoperators.
\bea
{}[h,{B}_n+\xi{\tilde B}_n]&=&{B}_n\left(-4\sin\frac{2\pi n}{M}
+4i\xi\left(1-\cos\frac{2\pi n}{M}\right)\right)\nonumber\\
&& +{\tilde B}_n\left(4\xi\sin\frac{2\pi n}{M}
-4i\left(1-\cos\frac{2\pi n}{M}\right)\right)\nonumber\\
&\equiv& \Delta({B}_n+\xi{\tilde B}_n).
\eea
This implies
\bea
1+\xi^2+2i\xi\cot\frac{\pi n}{M}&=&0.
\eea
Solving the quadratic gives
\bea
\xi_\pm&=&-i\cot\frac{\pi n}{M}\pm i\sqrt{1+\cot^2\frac{\pi n}{M}}
=\begin{cases}\displaystyle{i\tan\frac{\pi n}{2M}}\\
\phantom{.}\\
\displaystyle{-i\cot\frac{\pi n}{2M}}\end{cases}
\nonumber\\
\Delta_\pm&=&-4\sin\frac{2\pi n}{M}
+4i\xi_\pm\left(1-\cos\frac{2\pi n}{M}\right)=8\sin\frac{\pi n}{M}
\left(-\cos{\pi n}{M}+i\xi_\pm\sin{\pi n}{M}\right)\nonumber\\
&=&\mp8\sin\frac{\pi n}{M}.
\eea
We therefore define energy lowering operators,
\bea
F_n&=& B_n\cos\frac{\pi n}{2M}+i{\tilde B}_n\sin\frac{\pi n}{2M},
\eea
and raising operators,
\bea
{\bar F}_n&=&B_n\sin\frac{\pi n}{2M}-i{\tilde B}_n\cos\frac{\pi n}{2M},
\eea
which can be inverted
\bea
B_n&=&F_n\cos\frac{\pi n}{2M}+{\bar F}_n\sin\frac{\pi n}{2M},\qquad
i{\tilde B}=F_n\sin\frac{\pi n}{2M}-{\bar F}_n\cos\frac{\pi n}{2M}.
\eea
We notice that
\bea
F_n^\dagger&=&B_{M-n}\cos\frac{\pi n}{2M}-i{\tilde B}_{M-n}\sin\frac{\pi n}{2M}
={\bar F}_{M-n}\\
\{F_n,{\bar F}_m\}&=&2\sin\frac{\pi n}{2M}\cos\frac{\pi m}{2M}\delta_{n+m,M}
+2\cos\frac{\pi n}{2M}\sin\frac{\pi m}{2M}\delta_{n+m,M}=2\delta_{n+m,M}.
\eea
Applying $h$ to a state satisfying $F_n\ket{G}=0$ for all $n$, leads to a 
calculation of the ground energy. Remembering there is a contribution for
each of the $s$ components, we find
\bea
E_G&=&-4s\sum_{n=1}^{M-1}\sin\frac{n\pi}{M}\sim-\frac{8s}{\pi}M
+\frac{2\pi s}{3M}+O(M^{-3}).
\eea
Since $M=P^+/m$ is conserved in all processes, it can be harmlessly subtracted,
and we can identify the string tension by comparing the $1/M$ term to
the string $P^-_G$:
\bea
P^-_G=\frac{T_0}{4m}(E_G-8Ms/\pi)\sim \frac{\pi sT_0}{6P^+}(1+O(M^{-2})).
\eea
At $N=\infty$, the lowest squared mass in this model is $s\pi T_0/3>0$,
i.e. there is a mass gap.
\section{Three Closed Bit Chain Overlap}
As seen in Section 2, the terms in the $1/N$ expansion are determined 
by the overlap integrals $V_{rst}$ and $W_{rst}$. Let us focus on
the second of these. We can calculate it in the raising and lowering operator
formalism by expressing the ground state $\ket{G}$ of the large 
string in terms of raising and lowering operators of the two small strings,.
applied to the tensor product of the ground states of the small strings.

Divide the $M$ spin variables into $L$ ($k=1,\ldots L$) and $K=M-L$ 
($k=L+1,\ldots, M$) variables. Then for each subset we define modes
\bea
S_k&=& \frac{1}{\sqrt{L}}\sum_{n=0}^{L-1}B^{(1)}_n e^{2\pi ikn/L},\qquad
1\leq k\leq L\\
S_k&=& \frac{1}{\sqrt{K}}\sum_{n=0}^{K-1}B^{(2)}_n e^{2\pi i(k-L)n/(M-L)},
\qquad L+1\leq k\leq M,
\eea 
and likewise for ${\tilde S_k}$, putting a tilde on the corresponding
$B$'s. Then introduce the vectors
\bea
v_m^k&=&\frac{1}{\sqrt{M}}e^{2\pi ikm/M},\qquad k=1,\cdots,M;\quad m=0,\cdots, M-1\\
v^{(1)k}_n&=&\frac{1}{\sqrt{L}}e^{2\pi ikm/L},\qquad k=1,\cdots,L;\quad
n=0,\cdots, L-1\\
v^{(2)k}_n&=&\frac{1}{\sqrt{K}}e^{2\pi i(k-L)m/K},\qquad k=L+1,\cdots,M;\quad
n=0,\cdots, M-L-1.
\eea
Then the $B_n,{\tilde B}_n$ are related to the $B^{(1)}_n,{\tilde B}^{(1)}_n$
and $B^{(2)}_n,{\tilde B}^{(2)}_n$ by
\bea
B_m&=&\sum_{n=0}^{L-1}B_n^{(1)}v_m^\dagger v_n^{(1)}+\sum_{n=0}^{M-L-1}
B_n^{(2)}v_m^\dagger v^{(2)}_n\\
{\tilde B}_m&=&\sum_{n=0}^{L-1}{\tilde B}_n^{(1)}v_m^\dagger v_n^{(1)}+\sum_{n=0}^{M-L-1}
{\tilde B}_n^{(2)}v_m^\dagger v^{(2)}_n.
\label{overlapconditions}\eea
Zero modes require special attention. First we note that
\bea
B_0&=&B_0^{(1)}\sqrt{\frac{L}{M}}
+B_0^{(2)}\sqrt{\frac{K}{M}},\qquad {\tilde B}_0={\tilde B}_0^{(1)}\sqrt{\frac{L}{M}}
+{\tilde B}_0^{(2)}\sqrt{\frac{K}{M}}.\eea
It is then convenient to define the relative zero mode operators
\bea
b_0&=&B_0^{(1)}\sqrt{\frac{K}{M}}-B_0^{(2)}\sqrt{\frac{L}{M}},\qquad
{\tilde b}_0={\tilde B}_0^{(1)}\sqrt{\frac{K}{M}}-{\tilde B}_0^{(2)}\sqrt{\frac{L}{M}},
\eea
and it is easy to confirm the Clifford algebra
\bea
\{B_0,B_0\}&=&\{{\tilde B}_0,{\tilde B}_0\}=\{b_0,b_0\}
=\{{\tilde b}_0,{\tilde b}_0\}=2\nonumber\\
\{B_0,b_0\}&=&\{{\tilde B}_0,{\tilde b}_0\}=\{{B}_0,{\tilde b}_0\}
=\{{b}_0,{\tilde B}_0\}=\{{B}_0,{\tilde B}_0\}=\{{b}_0,{\tilde b}_0\}=0.
\eea
We can now rewrite the overlap conditions as
\bea
B_m&=&b_0v_m^\dagger w_0
+\sum_{n=1}^{L-1}B_n^{(1)}v_m^\dagger v_n^{(1)}+\sum_{n=1}^{M-L-1}
B_n^{(2)}v_m^\dagger v^{(2)}_n\\
{\tilde B}_m&=&{\tilde b}_0v_m^\dagger w_0
+\sum_{n=1}^{L-1}{\tilde B}_n^{(1)}v_m^\dagger v_n^{(1)}
+\sum_{n=1}^{M-L-1}
{\tilde B}_n^{(2)}v_m^\dagger v^{(2)}_n\\
w_0&\equiv&v_0^{(1)}\sqrt{\frac{K}{M}}-
v^{(2)}_0\sqrt{\frac{L}{M}}.
\eea
Finally, in order to calculate the 3 chain vertex, we relate the 
energy lowering operators for the large chain $F_m$ for $m\neq 0$ 
to the raising and lowering operators for the smaller chains. 
The zero mode operators
commute with $h$, but it is convenient to define 
$f_0=(b_0+i{\tilde b}_0)/{2}$ and ${\bar f}_0=f_0^\dagger
=(b_0-i{\tilde b}_0)/{2}$ which satisfy
\bea
f_0^2={\bar f}_0^2=0,\qquad \{f_0,{\bar f}_0\}=1.
\eea
Let $f_n$ be the $M-1$ operators $f_0,F^{(1)}_n/\sqrt{2},F^{(2)}_n/\sqrt{2}$, 
so that $\{f_m,f^\dagger_n\}=\delta_{mn}$, and we can write
\bea
F_m&=&\sqrt{2}\sum_{n=0}^{M-2}(f_nC_{mn}+f_n^\dagger S_{mn}),
\label{lowerlarge}
\eea
where the matrices $C, S$ are given in Appendix B.

Then we seek the ground state of the large chain in the form
\bea
\ket{G}&=&\exp\left\{\frac{1}{2}\sum_{k,l}M_{kl}f_k^\dagger f_l^\dagger\right\}
\ket{0}[{\det}(I+MM^\dagger)]^{-1/4},\eea
where $f_k\ket{0}=0$ for $k=0,\ldots,M-2$.
$F_m\ket{G}=0$ is equivalent to
\bea
C_{mn}M_{nl}+S_{ml}=0.
\eea
From $CM=-S$ we compute
\bea
C(I+MM^\dagger)C^\dagger&=&CC^\dagger+SS^\dagger\nonumber\\
\det C \det(I+MM^\dagger)\det C^\dagger&=&\det(CC^\dagger+SS^\dagger)\nonumber\\
\det(I+MM^\dagger)&=&\frac{\det(CC^\dagger+SS^\dagger)}{\det(CC^\dagger)}.
\eea
Using MATLAB to study these determinants numerically, we find that
$\det(CC^\dagger+SS^\dagger)=1$ and we also confirm the behavior
\bea
\det CC^\dagger&\sim&\frac{0.9290}{[KLM]^{1/6}}
\left(\frac{L}{M}\right)^{[M/K-L/M]/3-2/3}
\left(\frac{K}{M}\right)^{[M/L-K/M]/3-2/3}.
\eea
Here $K=M-L$ is the number of bits in one of the smaller strings.
It is interesting to compare this determinant for the Grassmann overlap
to the corresponding one for a single bosonic string coordinate,
\bea
\det XX^\dagger&=&\frac{2.1528}{[KLM]^{1/6}}
\left(\frac{L}{M}\right)^{-[M/K-L/M]/3}
\left(\frac{K}{M}\right)^{-[M/L-K/M]/3},
\eea
which was also calculated numerically with MATLAB.
This bosonic determinant, apart from the numerical factor, can be understood
based on the conformal mapping properties of the worldsheet
\cite{mandelstamlc,thornwsdet}.
There is an intimate relation between the Grassmann and bosonic 
overlaps reflected in the fact that
the superstring overlap involves the product of the two
\bea
\det CC^\dagger\det XX^\dagger&=&\frac{2.0000}{[KLM]^{1/3}}
\left(\frac{KL}{M^2}\right)^{-2/3}=2.0000\frac{M}{KL},
\eea
in which the combination is greatly simplified. This simplification is the
content of the Green-Schwarz statement that the bosonic and spinor worldsheet
determinants cancel each other; it is associated with the dependence
of an offshell vertex on the interaction time
$e^{-ia\Delta P^-}$. The measure contribution to $\Delta P^-$ is
\bea
\frac{s-d}{6}\Delta\frac{1}{2P^+}\to 0
\eea
for $d\to s$, which is the supersymmetry requirement. It is important
to appreciate that the cancellation is actually incomplete, and
moreover, the part left-over is essential to account for the
eventual Poincar\'e invariance.

Of course the string bit model studied in this article produces no bosonic 
coordinates but only Grassmann ones. As such, the requirement that
the vertex have a finite continuum limit, i.e. that it scale as
$M^{-3}$ at large $M$ with $L/M,K/M$ fixed, determines $s=24$. 
We call this interesting string model the ``protostring''.

\section{The Proto-String Theory}
To summarize our work, we have found that the Grassmann overlap scales
as $M^{-s/8}$ if there are $s$ Grassmann worldsheet fields. The scaling
of the bosonic overlap is $M^{-d/8}$ for $d$ transverse worldsheet scalars.
If one combines these one gets $M^{-(s+d)/8}$. With no operator
insertions at the break/join point, the
smooth continuum limit would require $s+d=24$. The bosonic string has
no Grassmann worldsheet fields so the critical dimension should
be $d=24$. The superstring has $s=d=8$ which does not give a smooth
continuum limit. But we also know that the superstring requires
an operator insertion proportional to $\Delta X^i\Delta X^j$ 
at the joining point.
This insertion produces an additional factor $M^{-1}$, which combined with
the $s=d=8$ overlaps ensures a smooth continuum limit. Poincar\'e
supersymmetry requires a further 8th order Grassmann polynomial $P_{ij}(S)$
which, as shown in Appendix C, has no effect on the overall scaling behavior.

The bosonic string has played an important role in the formal
string literature, because of the economy and simplicity of its
interactions--reflected in the absence of operator insertions.. 
We now see that there is another possibility that requires
no insertions. It is to have $s=24$ and $d=0$. This is a pure Grassmann analog
of the bosonic string and as such should be of some interest
in string theory. 
The superstring and the RNS string both require operator insertions
at the vertex break/join point. In order to compare the various possibilities
we need to know the scaling laws of various insertions. And for these we
need to know the overlap for the excited states. The matrix
$M$ is well-known in the continuum limit, and with that knowledge one
can obtain the needed scaling laws. In Appendix C we discuss these
issues for insertions of $S$ variables, with the conclusion that they
scale as $M^0$. To compare to the other major possibilities we
have prepared tables of the various scaling laws.

We first note the nominal scaling rules for insertions. For bosonic variables
the insertion  $\Delta X=X_{k+1}-X_k\sim m \frac{\partial X}{\partial\sigma}$
nominally scales as $\Delta X\sim M^{-1}$.
Similarly $S_k,\Gamma_k\sim \sqrt{m}(S(\sigma),\Gamma(\sigma))$ 
nominally scales as $S_k,\Gamma_k\sim M^{-1/2}$. Here $\Gamma^k$ is
an RNS fermionic worldsheet field. However the
fission/fusion singularity enhances these expectations:as illustrated
in Table~\ref{enhancements}.
\begin{table}[ht]
\begin{center}
\begin{tabular}{|c|c|c|}
\hline 
Insertion&Enhancement&Net\\
\hline
$\Delta X$& $M^{1/2}$& $M^{-1/2}$\\
$ S$& $M^{1/2}$& $M^0$\\
$\Gamma$& $M^{1/4}$&$M^{-1/4}$\\
\hline
\end{tabular} 
\caption{Enhancement of scaling laws for operator insertions on
overlaps of the bosonic ($\Delta X$), Green-Schwarz ($S$), and RNS ($\Gamma$)
types.}
\label{enhancements}
\end{center}
\end{table}
Note that the enhancement is different in the RNS and Green-Schwarz 
overlaps. The various overlap scaling laws are compared in Table
\ref{overlaps}.
\begin{table}[ht]
\begin{center}
\begin{tabular}{|c|c||c|c|}
\hline
Overlap& Scaling & Insertion& Scaling\\ 
\hline
$V_X$ &  $M^{-d/8}$ &    $\Delta X$& $M^{-1/2}$\\
$V_S$ &  $M^{-s/8}$&  $S$& $M^0$\\
$V_{\Gamma}$  & $M^{-d/16}$ & $\Gamma$ & $M^{-1/4}$
\\
\hline
\end{tabular}
\caption{Summary of overlap scaling laws along with insertion rules.}
\label{overlaps}
\end{center}
\end{table}
Finally, the various string models with total vertices and 
the critical dimension,
determined by requiring the total vertex to scale as $M^{-3}$,
are displayed in Table~\ref{criticald}.
\begin{table}[ht]
\begin{center}
\begin{tabular}{|c||c|c|}
\hline
Type & Total Vertex & Critical Dimension\\
\hline 
Bosonic String& $V_X$& $d=24$\\
IIB Superstring&     $\Delta X^i\Delta X^j{\cal P}_{ij}(S)V_XV_S$ &$d=s=8$\\
RNS String &    $(\Gamma\cdot\Delta X)^2V_\Gamma V_X$& $d=8$\\
Protostring&   $V_S$& $s=24$\\
\hline
\end{tabular}
\caption{Total Vertices and critical dimensions for bosonic, Green-Schwarz,
RNS, and Protostring overlaps}
\label{criticald}
\end{center}
\end{table}
As we have discussed, the protostring is a 
Grassmann analog of the bosonic string, in that
neither require operator insertions. However, there are striking differences.
For one, the bosonic string has a tachyonic ground state, whereas the
lowest mass squared of the protostring is positive. Accordingly, the
protostring is stable. Another interesting feature of the protostring is 
that its worldsheet
degrees of freedom match those of the superstring: 16 of the Grassmann
worldsheet fields can be bosonized into 8 compactified bosonic worldsheet 
fields. These, together with the remaining 8 Grassmann fields,
match the worldsheet
fields of the superstring. However, the compactification radius
is fixed, so it is not obvious how to achieve large spatial dimensions. 
There is the hope that some deformation of the protostring,
which enables large transverse dimensions, can
be found to produce the actual superstring. Taken as it is given here, the
protostring moves in 1 space dimension (there are no large 
transverse dimensions).  The stability of the protostring recommends it as
a solid starting point for defining string theory more generally..Exploring
this possibility is a promising directioin for future research. 
\vskip14pt
\noindent\underline{Acknowledgments}: I would like to thank the organizers
of the Miami 2015 Conference on Physics where these results
were first reported. In particular, I thank
Pierre Ramond, Arkady Tseytlin,
Phillip Mannheim, and Ethan Torres for
insightful questions and helpful comments. 
This research was supported in part by the Department
of Energy under Grant No. DE-SC0010296. 
\appendix
\section{The Hamiltonian and its action on color singlets}
The string bit creation and annihilation operators satisfy the
(anti)-commutation relations
\bea
{}[(\phi_{a_1\cdots a_n})_\alpha^{\ \beta},({\bar\phi}_{b_1\cdots b_n})_\gamma^{\ \delta}]_\pm&\equiv&(\phi_{a_1\cdots a_n})_\alpha^{\ \beta}({\bar\phi}_{b_1\cdots b_n})_\gamma^{\ \delta}-(-)^{mn}({\bar\phi}_{b_1\cdots b_n})_\gamma^{\ \delta}(\phi_{a_1\cdots a_n})_\alpha^{\ \beta}\nonumber\\
&=&\delta_{mn}\delta_{\alpha}^{\ \delta}\delta_{\gamma}^{\ \beta}\sum_P(-)^P\delta_{a_1b_{P_1}}\cdots\delta_{a_nb_{P_n}},
\label{crs}
\eea
which incorporate the fact that ${\bar\phi}$ creates a boson if $n$ is
even and a fermion if $n$ is odd. The sum over $P$ is over all permutation
of $1,2,\ldots,n$.

The Hamiltonian analyzed in this paper is the one called $H_S$ in
\cite{thornspace}. We quote
\bea
H_S&=& H_1+H_2+H_3+H_4+H_5,
\eea
where the $H_i$ are:
\bea
H_1&=&\frac{2}{N}\sum_{n=0}^s\sum_{k=0}^s\frac{s-2n}{n!k!}
\Tr{\bar\phi}_{a_1\cdots a_n}{\bar\phi}_{b_1\cdots b_k}
{\phi}_{b_1\cdots b_k}{\phi}_{a_1\cdots a_n}\\
H_2&=&\frac{2}{N}\sum_{n=0}^{s-1}\sum_{k=0}^{s-1}\frac{(-)^k}{n!k!}
\Tr{\bar\phi}_{a_1\cdots a_n}{\bar\phi}_{bb_1\cdots b_k}
{\phi}_{b_1\cdots b_k}{\phi}_{ba_1\cdots a_n}\\
H_3&=&\frac{2}{N}\sum_{n=0}^{s-1}\sum_{k=0}^{s-1}\frac{(-)^k}{n!k!}
\Tr{\bar\phi}_{ba_1\cdots a_n}{\bar\phi}_{b_1\cdots b_k}
{\phi}_{bb_1\cdots b_k}{\phi}_{a_1\cdots a_n}\\
H_4&=&\frac{2i}{N}\sum_{n=0}^{s-1}\sum_{k=0}^{s-1}\frac{(-)^k}{n!k!}
\Tr{\bar\phi}_{a_1\cdots a_n}{\bar\phi}_{b_1\cdots b_k}
{\phi}_{bb_1\cdots b_k}{\phi}_{ba_1\cdots a_n}\\
H_5&=&-\frac{2i}{N}\sum_{n=0}^{s-1}\sum_{k=0}^{s-1}\frac{(-)^k}{n!k!}
\Tr{\bar\phi}_{ba_1\cdots a_n}{\bar\phi}_{bb_1\cdots b_k}
{\phi}_{b_1\cdots b_k}{\phi}_{a_1\cdots a_n}.
\label{sham}
\eea
$H_S$ commutes with the supersymmetry operators
\bea
Q^a&=&\sum_{n=0}^{s-1}\frac{(-)^n}{n!}\Tr\left[e^{i\pi/4}
{\bar\phi}_{a_1\cdots a_n}\phi_{aa_1\cdots a_n}+e^{-i\pi/4}
{\bar\phi}_{aa_1\cdots a_n}\phi_{a_1\cdots a_n}\right]\nonumber\\
\{Q^a,Q^b\}&=&2M\delta_{ab},
\eea
which will guarantee equal numbers of bosonic and fermionic eigenstates
at each energy level.

Using the commutation relations (\ref{crs}), it is straightforward 
to obtain the action of the $H_i$ on single trace states:
\bea
H_1T(\theta_1,\cdots,\theta_M)\ket{0}&=&2\sum_{k=1}^M\left(s-2\theta_k^a
\frac{d}{d\theta_k^a}\right)
T(\theta_1,\cdots,\theta_M)\ket{0}\nonumber\\
&&\hskip-.25in +\frac{2}{N}\sum_{k=1}^M\left(s-2\theta_k^a
\frac{d}{d\theta_k^a}\right)\sum_{l\neq k,k+1}T(\theta_l\cdots\theta_k)
T(\theta_{k+1}\cdots\theta_{l-1})\ket{0}
\eea
\bea
H_2T(\theta_1,\cdots,\theta_M)\ket{0}&=&2\sum_{k=1}^M
\theta_k^a\frac{d}{d\theta_{k+1}^a}T(\theta_1,\cdots,\theta_M)\ket{0}\nonumber\\
&&\hskip-.25in +\frac{2}{N}\sum_{k=1}^M\sum_{l\neq k,k+1}
\theta_k^a\frac{d}{d\theta_{l}^a}T(\theta_l\cdots\theta_k)
T(\theta_{k+1}\cdots\theta_{l-1})\ket{0}
\eea
\bea
H_3T(\theta_1,\cdots,\theta_M)\ket{0}&=&2\sum_{k=1}^M
\theta_{k+1}^a\frac{d}{d\theta_{k}^a}T(\theta_1,\cdots,\theta_M)\ket{0}\nonumber\\
&&\hskip-.25in +\frac{2}{N}\sum_{k=1}^M\sum_{l\neq k,k+1}
\theta_l^a\frac{d}{d\theta_{k}^a}T(\theta_l\cdots\theta_k)
T(\theta_{k+1}\cdots\theta_{l-1})\ket{0}
\eea
\bea
H_4T(\theta_1,\cdots,\theta_M)\ket{0}&=&-2i\sum_{k=1}^M
\theta_k^a\theta^a_{k+1}T(\theta_1,\cdots,\theta_M)\ket{0}\nonumber\\
&&\hskip-.25in -\frac{2i}{N}\sum_{k=1}^M\sum_{l\neq k,k+1}
\theta_k^a{\theta_{l}^a}T(\theta_l\cdots\theta_k)
T(\theta_{k+1}\cdots\theta_{l-1})\ket{0}
\\
H_5T(\theta_1,\cdots,\theta_M)\ket{0}&=&-2i\sum_{k=1}^M
\frac{d}{d\theta_k^a}\frac{d}{d\theta^a_{k+1}}T(\theta_1,\cdots,\theta_M)\ket{0}
\nonumber\\
&&\hskip-.25in -\frac{2i}{N}\sum_{k=1}^M\sum_{l\neq k,k+1}
\frac{d}{d\theta_k^a}\frac{d}{d\theta_{l}^a}T(\theta_l\cdots\theta_k)
T(\theta_{k+1}\cdots\theta_{l-1})\ket{0}.\eea
We note that the differential operators are applied to nearest neighbors on the
same trace when they involve two distinct Grassmann variables.

The action of the $H_i$ on multitrace states takes two forms. 
When both annihilation operators contract on the same trace,
 the action can be read off from the
preceding formulas. When they act on different traces the action is to
fuse them into a single trace as follows
\bea
H_1T({\theta_1\cdots\theta_K})
T({\eta_{1}\cdots\eta_L})\ket{0}_{\rm Fusion}&=&
\nonumber\\
&&\hskip-2.5in +\frac{2}{N}\sum_{k=1}^K\sum_{l=1}^L \left(s-2\theta_k^a
\frac{d}{d\theta_k^a}\right)T({\theta_{k+1}
\cdots\theta_{k}}\eta_l\cdots\eta_{l-1})\ket{0}
\nonumber\\
&&\hskip-2.5in +\frac{2}{N}\sum_{k=1}^K\sum_{l=1}^L  \left(s-2\eta_l^a
\frac{d}{d\eta_l^a}\right)T(\theta_k\cdots\theta_{k-1}{\eta_{l+1}
\cdots\eta_{l}})\ket{0}\\
H_2T({\theta_1\cdots\theta_K})
T({\eta_{1}\cdots\eta_L})\ket{0}_{\rm Fusion}&=&
\nonumber\\
&&\hskip-2.5in +\frac{2}{N}\sum_{k=1}^K\sum_{l=1}^L \theta_k^a\frac{d}{d\eta_l}T({\theta_{k+1}
\cdots\theta_{k}}\eta_l\cdots\eta_{l-1})\ket{0}
\nonumber\\
&&\hskip-2.5in +\frac{2}{N}\sum_{k=1}^K\sum_{l=1}^L  
\eta_l^a\frac{d}{d\theta_k}T(\theta_k\cdots\theta_{k-1}{\eta_{l+1}
\cdots\eta_{l}})\ket{0}.\label{fuse1}
\eea
with similar transcriptions for the other $H_i$. In each case the differential
operators have the same structure as the fission terms, but the states on the
right are a suitable pair of single trace states. And when there are two 
distinct Grassmann operators they act on nearest neighbors on the large trace.
\section{Formulas for Overlap Calculations}
The following matrix elements are needed in (\ref{overlapconditions}).
\bea
v^\dagger_m v^{(1)}_n&=&\frac{1}{\sqrt{ML}}\sum_{k=1}^Le^{2i\pi k(n/L-m/M)}
=-\frac{1}{\sqrt{ML}}\frac{1-e^{-2\pi imL/M}}{1-e^{-2i\pi (n/L-m/M)}}\\
v^\dagger_m v^{(2)}_n&=&\frac{1}{\sqrt{MK}}\sum_{k=L+1}^M
e^{2i\pi [(k-L)(n/K-m/M)-Lm/M]}\nonumber\\
&=&\frac{1}{\sqrt{MK}}
\frac{1-e^{-2\pi imL/M}}{1-e^{-2i\pi (n/K-m/M)}}\\
v^\dagger_m w_0&=&\frac{-1}{\sqrt{ML}}\frac{1-e^{-2\pi imL/M}}{1-e^{2i\pi m/M}}
\sqrt{\frac{K}{M}}-\frac{1}{\sqrt{MK}}
\frac{1-e^{-2\pi imL/M}}{1-e^{2i\pi m/M)}}\sqrt{\frac{L}{M}}\nonumber\\
&=&-\frac{1}{\sqrt{LK}}\frac{1-e^{-2\pi imL/M}}{1-e^{2i\pi m/M}}.\eea
Then
\bea
F_m&=&\sum_{n=1}^{L-1}\left[F^{(1)}_n\cos\left(\frac{n\pi}{2L}-\frac{m\pi}{2M}\right)+{\bar F}^{(1)}_n\sin\left(\frac{n\pi}{2L}-\frac{m\pi}{2M}\right)\right]
v_m^\dagger v^{(1)}_n\nonumber\\
&&+\sum_{n=1}^{M-L-1}\left[F^{(2)}_n\cos\left(\frac{n\pi}{2K}
-\frac{m\pi}{2M}\right)+{\bar F}^{(2)}_n\sin\left(\frac{n\pi}{2K}
-\frac{m\pi}{2M}\right)\right]v_m^\dagger v^{(2)}_n\nonumber\\
&&+\left(f_0\cos\left(\frac{m\pi}{2M}-\frac{\pi}{4}\right)
+{\bar f}_0\cos\left(\frac{m\pi}{2M}+\frac{\pi}{4}\right)\right)
v_m^\dagger w_0\sqrt{2}\nonumber\\
&=&\sum_{n=1}^{L-1}\left[F^{(1)}_n\cos\left(\frac{n\pi}{2L}-\frac{m\pi}{2M}\right)v_m^\dagger v^{(1)}_n+{F}^{(1)\dagger}_n\cos\left(\frac{n\pi}{2L}+\frac{m\pi}{2M}\right)v_m^\dagger v^{(1)}_{L-n}\right]
\nonumber\\
&&+\sum_{n=1}^{M-L-1}\left[F^{(2)}_n\cos\left(\frac{n\pi}{2K}
-\frac{m\pi}{2M}\right)v_m^\dagger v^{(2)}_n
+{F}^{(2)^\dagger}_n\cos\left(\frac{n\pi}{2K}
+\frac{m\pi}{2M}\right)v_m^\dagger v^{(2)}_{M-L-n}\right]\nonumber\\
&&+\left(f_0\cos\left(\frac{m\pi}{2M}-\frac{\pi}{4}\right)
+{f}_0^\dagger\cos\left(\frac{m\pi}{2M}+\frac{\pi}{4}\right)\right)
v_m^\dagger w_0\sqrt{2}.
\eea
Then the $C,S$ matrices needed in (\ref{lowerlarge}) are given by:
\bea
C_{m0}&=&-\frac{1}{\sqrt{LK}}\frac{1-e^{-2\pi imL/M}}{1-e^{2i\pi m/M}}\cos\left(\frac{m\pi}{2M}-\frac{\pi}{4}\right)\nonumber\\
C_{mn1}&=&-\frac{1}{\sqrt{ML}}\frac{1-e^{-2\pi imL/M}}{1-e^{-2i\pi (n/L-m/M)}}\cos\left(\frac{n\pi}{2L}-\frac{m\pi}{2M}\right)\nonumber\\
C_{mn2}&=&\frac{1}{\sqrt{MK}}
\frac{1-e^{-2\pi imL/M}}{1-e^{-2i\pi (n/K-m/M)}}\cos\left(\frac{n\pi}{2K}
-\frac{m\pi}{2M}\right)\\
S_{m0}&=&-\frac{1}{\sqrt{LK}}\frac{1-e^{-2\pi imL/M}}{1-e^{2i\pi m/M}}\cos\left(\frac{m\pi}{2M}+\frac{\pi}{4}\right)\nonumber\\
S_{mn1}&=&-\frac{1}{\sqrt{ML}}\frac{1-e^{-2\pi imL/M}}{1-e^{2i\pi (n/L+m/M)}}
\cos\left(\frac{n\pi}{2L}+\frac{m\pi}{2M}\right)\nonumber\\
S_{mn2}&=&\frac{1}{\sqrt{MK}}
\frac{1-e^{-2\pi imL/M}}{1-e^{2i\pi (n/K+m/M)}}\cos\left(\frac{n\pi}{2K}
+\frac{m\pi}{2M}\right).\eea

\section{Constructing $\ket{G}$ in the continuum limit}
The equations determining the matrix $M$ can be analyzed in the continuum limit
in which $L,M\to\infty$ with $x\equiv L/M$ fixed. Then $K/M\equiv
(M-L)/M=1-x$. For this purpose
we consider this limit on the matrices $C,S$. This limit must
be taken in eight separate cases corresponding to left and right moving
waves on each of the three closed strings. It is convenient to remove
some common factors of $C,S$ using lower case letters for the reduced matrices:
\bea
C_{mn}\equiv \frac{1-e^{-2\pi imL/M}}{2\pi i}c_{mn},\qquad 
S_{mn}\equiv \frac{1-e^{-2\pi imL/M}}{2\pi i}s_{mn}.
\eea 
Then
\begin{enumerate}
\item Holding $m,n1,n2$ fixed
\bea
c_{m0}&\to&\frac{1}{m\sqrt{2x(1-x)}}\nonumber\\
c_{mn1}&\to&\frac{-1}{{n/\sqrt{x}-m\sqrt{x}}},\qquad
c_{mn2}\to\frac{1}{{ n/\sqrt{1-x}-m\sqrt{1-x}}}\\
s_{m0}&\to&\frac{1}{m\sqrt{2x(1-x)}}
\nonumber\\
s_{mn1}&\to&\frac{1}{{n/\sqrt{x}+m\sqrt{x}}},\qquad
s_{mn2}\to\frac{-1}{{n/\sqrt{1-x}+m\sqrt{1-x}}},
\eea
\item Holding $m^\prime=M-m,n1,n2$ fixed
\bea
c_{m0}&\to&\frac{-1}{ m^\prime \sqrt{2x(1-x)}},\qquad
c_{mn1}\to0,\qquad c_{mn2}\to 0\\
s_{m0}&\to&\frac{1}{ m^\prime \sqrt{2x(1-x)}},\qquad
s_{mn1}\to0,\qquad s_{mn2}\to0,
\eea
\item Holding $m,n^\prime1\equiv L-n1,n2$ fixed
\bea
c_{m0}&=&\frac{1}{m\sqrt{2x(1-x)}},\qquad
c_{mn1}=0,\qquad
c_{mn2}=\frac{1}{{ n/\sqrt{1-x}-m\sqrt{1-x}}}
\\
s_{m0}&=&\frac{1}{m\sqrt{2x(1-x)}},\qquad
s_{mn1}=0,\qquad
s_{mn2}=\frac{-1}{{n/\sqrt{1-x}+m\sqrt{1-x}}},\eea
\item Holding $m^\prime,n^\prime1,n2$ fixed
\bea
c_{m0}&=&\frac{-1}{ m^\prime \sqrt{2x(1-x)}},\qquad
c_{mn1}=\frac{1}{ (n^\prime /\sqrt{x}-m^\prime\sqrt{x})},\qquad
c_{mn2}=0\\
s_{m0}&=&\frac{1}{ m^\prime \sqrt{2x(1-x)}},\qquad
s_{mn1}=\frac{1}{(n^\prime /\sqrt{x}+m^\prime\sqrt{x})},\qquad
s_{mn2}=0,\eea
\item Holding $m,n1,n^\prime2=K-n2$ fixed
\bea
c_{m0}&=&\frac{1}{m\sqrt{2x(1-x)}},\qquad
c_{mn1}=\frac{-1}{{n/\sqrt{x}-m\sqrt{x}}},\qquad
c_{mn2}=0\\
s_{m0}&=&\frac{1}{m\sqrt{2x(1-x)}},\qquad
s_{mn1}=\frac{1}{{n/\sqrt{x}+m\sqrt{x}}},\qquad
s_{mn2}=0,\eea
\item Holding $m^\prime,n1,n^\prime2$ fixed
\bea
c_{m0}&=&\frac{-1}{ m^\prime \sqrt{2x(1-x)}},\qquad
c_{mn1}=0,\qquad
c_{mn2}=\frac{-1}{(n^\prime /\sqrt{1-x}-m^\prime\sqrt{1-x})}\\
s_{m0}&=&\frac{1}{ m^\prime \sqrt{2x(1-x)}},\qquad
s_{mn1}=0,\qquad
s_{mn2}=\frac{1}{(n^\prime /\sqrt{1-x}+m^\prime\sqrt{1-x})},\eea
\item Holding $m,n^\prime1,n^\prime2$ fixed
\bea
c_{m0}&=&\frac{1}{m\sqrt{2x(1-x)}},\qquad
c_{mn1}=0,\qquad c_{mn2}=0\\
s_{m0}&=&\frac{1}{m\sqrt{2x(1-x)}},\qquad
s_{mn1}=0,\qquad s_{mn2}=0,\eea
\item Holding $m^\prime,n^\prime1,n^\prime2$ fixed
\bea
c_{m0}&=&\frac{-1}{ m^\prime \sqrt{2x(1-x)}}
\nonumber\\
c_{mn1}&=&\frac{1}{ (n^\prime /\sqrt{x}-m^\prime\sqrt{x})},\qquad
c_{mn2}=
\frac{-1}{(n^\prime /\sqrt{1-x}-m^\prime\sqrt{1-x})}\\
s_{m0}&=&\frac{1}{ m^\prime \sqrt{2x(1-x)}}
\nonumber\\
s_{mn1}&=&\frac{1}{(n^\prime /\sqrt{x}+m^\prime\sqrt{x})},\qquad
s_{mn2}=
\frac{-1}{(n^\prime /\sqrt{1-x}+m^\prime\sqrt{1-x})}.\eea
\end{enumerate}
The equation $CM+S=0$ then breaks up into the series of equations
\bea
C_{m0}M_{0,l}+C_{mn1}M_{n1,l}+C_{mn2}M_{n2,l}+S_{ml}&=&0\label{first}\\
C_{mn1}M_{n1,0}+C_{mn2}M_{n2,0}+S_{m0}&=&0\label{second}\\
C_{m^\prime0}M_{0,l^\prime}+C_{m^\prime n1^\prime}M_{n1^\prime,l^\prime}
+C_{m^\prime n2^\prime}M_{n2^\prime,l^\prime}+S_{m^\prime,l^\prime}&=&0
\label{third}\\
C_{m^\prime n1^\prime}M_{n1^\prime,0}+C_{m^\prime n2^\prime}M_{n2^\prime,0}
+S_{m^\prime0}&=&0\label{fourth}\\
C_{m0}M_{0,l^\prime}+C_{mn1}M_{n1,l^\prime}+C_{mn2}M_{n2,l^\prime}&=&0
\label{fifth}\\
C_{m^\prime0}M_{0,l}+C_{m^\prime n1^\prime}M_{n1^\prime,l}+C_{m^\prime n2^\prime}M_{n2^\prime,l}&=&0,\label{sixth}
\eea
where unprimed indices refer to the continuum limit holding $m$, $n1$, or $n2$,
fixed; and primed indices indicate holding $M-m$, $L-n1$, or $M-L-n2$ fixed.
In these formulas $l$ is allowed to refer to either of the smaller
strings, so unprimed it is held fixed and primed $L-l$ or $M-L-l$
as appropriate is held fixed. In addition to these equations the
matrix $M$ is required to be ant-symmetric $M^T=-M$.
\subsection{Solving the continuum equations}
We can solve these equations using a method due to J. Goldstone,
who solved the analogous equations for the three bosonic open string
vertex \cite{goldstone}. The final results can also be found in
\cite{greenschwarzbrink}, which employs a different
method. Since the matrices $C,S$ involve reciprocals of linear combinations
of integers, one guesses a function with poles at appropriate points.
Goldstone's choice was
\bea
g(z)&=&\frac{\Gamma(1+zx)\Gamma(1+z(1-x))}{z\Gamma(1+z)}\frac{e^{z\xi}}{\sqrt{x(1-x)}}\\
\xi&\equiv& -x\ln(x)-(1-x)\ln(1-x).\eea
The function $g(z)$ has poles at $0$, $-n/x$, and $-n/(1-x)$ for $n$ positive
integers. 
\bea
g(z)&\sim& \frac{1}{ng(n/x)}\frac{1}{z+n/x},\qquad g(z)\sim 
\frac{1}{ng(n/(1-x))}\frac{1}{z+n/(1-x)},
\eea
respectively. At large $z$, $g$ behaves as $\sqrt{2\pi}z^{-1/2}$. 
Since $g(z)$ has zeroes
at $-1,-2,-3,\ldots$, $g(z)/(z+m)$ has the same poles as $g$ as long as $m$
is a positive integer. Then we can we can expand
\bea
\frac{g(z)}{z+m}&=&\sum_{n=1}^\infty \frac{1}{ng(n/x)}\frac{1}{z+n/x}
\frac{1}{m-n/x}+\sum_{n=1}^\infty \frac{1}{ng(n/(1-x))}\frac{1}{z+n/(1-x)}
\frac{1}{m-n/(1-x)}\nonumber\\
&&+\frac{1}{mz\sqrt{x(1-x)}}\nonumber\\
&=&\sum_{n=1}^\infty \frac{1}{ng(n/x)}\frac{1}{z+n/x}
\sqrt{x}c_{mn1}-\sum_{n=1}^\infty \frac{1}{ng(n/(1-x))}\frac{1}{z+n/(1-x)}
\sqrt{1-x}c_{mn2}\nonumber\\
&&+\frac{\sqrt{2}}{z}c_{m0}.
\label{id1}\eea
We can recognize this as the first of our equations to solve if we put
$z=l/x$ or $z=l/(1-x)$. Then the left side becomes either
$g(l/x)\sqrt{x}s_{ml1}$ or $g(l/(1-x))\sqrt{1-x}s_{ml2}$:
\bea
s_{ml1}&=&\sum_{n=1}^\infty \frac{1}{ng(n/x)g(l/x)}\frac{1}{l/x+n/x}
c_{mn1}\nonumber\\
&&-\sum_{n=1}^\infty \frac{1}{ng(n/(1-x))g(l/x)}\frac{1}{l/x+n/(1-x)}\sqrt{\frac{1-x}{x}}c_{mn2}+\frac{\sqrt{2x}}{g(l/z)l}c_{m0}\label{s1a}\\
-s_{ml2}&=&\sum_{n=1}^\infty \frac{1}{ng(n/x)g(l/(1-x)}\frac{1}{l/(1-x)+n/x}
\sqrt{\frac{x}{1-x}}c_{mn1}\nonumber\\
&&\hskip-.5in-\sum_{n=1}^\infty \frac{1}{ng(n/(1-x))g(l/(1-x)}\frac{1}{l/(1-x)+n/(1-x)}c_{mn2}+\frac{\sqrt{2(1-x)}}{g(l/(1-x))l}c_{m0}.
\label{s2a}\eea 
Unfortunately,
the inferred $M_{nl}$ would not be antisymmetric. We can fix this by noticing
the identity obtained by expanding $zg(z)/(z+m)$
\bea
\frac{zg(z)}{z+m}
&=&\sum_{n=1}^\infty \frac{-n/x}{ng(n/x)}\frac{1}{z+n/x}
\sqrt{x}c_{mn1}\nonumber\\
&&-\sum_{n=1}^\infty \frac{-n/(1-x)}{ng(n/(1-x))}\frac{1}{z+n/(1-x)}\sqrt{1-x}c_{mn2}.
\label{id2}
\eea
Putting $z=l/x$ and $z=l/(1-x)$ gives
\bea
s_{ml1}&=&
\sum_{n=1}^\infty \frac{-1}{lg(l/x)g(n/x)}\frac{1}{l/x+n/x}
c_{mn1}\nonumber\\
&&-\sum_{n=1}^\infty \frac{-1}{lg(l/x)g(n/(1-x))}\frac{1}{l/x+n/(1-x)}\sqrt{\frac{x}{1-x}}c_{mn2}\label{s1b}\\
-s_{ml2}&=&\sum_{n=1}^\infty \frac{-1}{lg(l/(1-x))g(n/x)}\frac{1}{l/(1-x)+n/x}
\sqrt{\frac{1-x}{x}}c_{mn1}\nonumber\\
&&-\sum_{n=1}^\infty \frac{-1}{lg(l/(1-x))g(n/(1-x))}\frac{1}{z+n/(1-x)}c_{mn2}.
\label{s2b}\eea
Taking the average of the two expressions for $s_{ml}$ gives an antisymmetric 
solution to the first equation
\bea
s_{ml1}&=&\frac{1}{2}\sum_{n=1}^\infty \frac{l-n}{lng(n/x)g(l/x)}\frac{1}{l/x+n/x}
c_{mn1}\nonumber\\
&&-\frac{1}{2}\sum_{n=1}^\infty \frac{l\sqrt{(1-x)/x}-n\sqrt{x/(1-x)}}{lng(n/(1-x))g(l/x)}\frac{1}{l/x+n/(1-x)}c_{mn2}+\frac{\sqrt{x/2}}{g(l/z)l}c_{m0}\\
-s_{ml2}&=&\frac{1}{2}\sum_{n=1}^\infty \frac{l\sqrt{x/(1-x)}-n\sqrt{(1-x)/x}}{ng(n/x)g(l/(1-x)}\frac{1}{l/(1-x)+n/x}c_{mn1}\nonumber\\
&&\hskip-.5in-\frac{1}{2}\sum_{n=1}^\infty \frac{l-n}{lng(n/(1-x))g(l/(1-x)}\frac{1}{l/(1-x)+n/(1-x)}c_{mn2}+\frac{\sqrt{(1-x)/2}}{g(l/(1-x))l}c_{m0}.\eea
From these equations we can read off some of the matrix elements of $M$:
\bea
M_{n1,l1}&=&-\frac{1}{2}\frac{l-n}{lng(n/x)g(l/x)}\frac{1}{l/x+n/x}\\
M_{n2,l1}&=&\frac{1}{2}\frac{l\sqrt{(1-x)/x}-n\sqrt{x/(1-x)}}{lng(n/(1-x))g(l/x)}\frac{1}{l/x+n/(1-x)}\\
M_{0,l1}&=&-\frac{\sqrt{x/2}}{g(l/x)l}\\
M_{n1,l2}&=&\frac{1}{2}\frac{l\sqrt{x/(1-x)}-n\sqrt{(1-x)/x}}{ng(n/x)g(l/(1-x)}\frac{1}{l/(1-x)+n/x}\\
M_{n2,l2}&=&-\frac{1}{2}\frac{l-n}{lng(n/(1-x))g(l/(1-x)}\frac{1}{l/(1-x)+n/(1-x)}\\
M_{0,l2}&=&\frac{\sqrt{(1-x)/2}}{g(l/(1-x))l}.
\eea
Next let's consider the equation (\ref{second}). We examine (\ref{id1})
and (\ref{id2}) near $z=0$. First put $z=0$ in (\ref{id2}).
\bea
\frac{1}{m\sqrt{x(1-x)}}=\sqrt{2}s_{m0}
&=&\sum_{n=1}^\infty \frac{-1}{ng(n/x)}
\sqrt{x}c_{mn1}\nonumber\\
&&-\sum_{n=1}^\infty \frac{-1}{ng(n/(1-x))}\sqrt{1-x}c_{mn2},
\eea
from which we confirm that $M_{n1,0}=-M_{0,n1}$ and $M_{n2,0}=-M_{0,n2}$.
As it turns out we don't need (\ref{id1}) here.

The analysis of equations (\ref{third}) and (\ref{fourth}) parallels that
of (\ref{first}) and (\ref{second}). Indeed, inspection of the 
relevant equations shows that $M_{n^\prime,l^\prime}=-M_{nl}$. It then remains to
analyze the first two equations (\ref{fifth}) and (\ref{sixth}), which
do not involve the $s$'s. We therefore examine the difference between the 
equations (\ref{s1a}), (\ref{s2a}) and (\ref{s1b}), (\ref{s2b}): 
\bea
0&=&\sum_{n=1}^\infty \frac{x}{lng(n/x)g(l/x)}
c_{mn1}\nonumber\\
&&-\sum_{n=1}^\infty \frac{\sqrt{x(1-x)}}{lng(n/(1-x))g(l/x)}
c_{mn2}+\frac{\sqrt{2x}}{g(l/z)l}c_{m0}\label{zero1}\\
0&=&\sum_{n=1}^\infty \frac{\sqrt{x(1-x)}}{lng(n/x)g(l/(1-x)}c_{mn1}\nonumber\\
&&\hskip-.5in-\sum_{n=1}^\infty \frac{1-x}{lng(n/(1-x))g(l/(1-x)}
c_{mn2}+\frac{\sqrt{2(1-x)}}{g(l/(1-x))l}c_{m0}.
\label{zero2}\eea 
The $c$'s in (\ref{sixth}) are the negatives of those in (\ref{fifth}),
but since they are homogeneous in $c$ the two equations are actually
identical in form. The coefficient of $c_{m0}$ in (\ref{zero1}) is
seen to be $-2M_{0,l1}=2M_{0,l1^\prime}$ and in (\ref{zero2}) to be 
$+2M_{0,l2}=-2M_{0,l2^\prime}$. We thus determine from (\ref{fifth})
\bea
M_{n1,l1^\prime}&=&\frac{x}{2lng(n/x)g(l/x)},\qquad 
M_{n2,l1^\prime}=-\frac{\sqrt{x(1-x)}}{2lng(n/(1-x))g(l/x)}\\
M_{n2,l2^\prime}&=&\frac{1-x}{2lng(n/(1-x))g(l/(1-x)},\qquad
M_{n1,l2^\prime}=-\frac{\sqrt{x(1-x)}}{2lng(n/x)g(l/(1-x)},
\eea
and from (\ref{sixth})
\bea
M_{n1^\prime,l1}&=&-\frac{x}{2lng(n/x)g(l/x)},\qquad 
M_{n2^\prime,l1}=\frac{\sqrt{x(1-x)}}{2lng(n/(1-x))g(l/x)}\\
M_{n2^\prime,l2}&=&-\frac{1-x}{2lng(n/(1-x))g(l/(1-x)},\qquad
M_{n1^\prime,l2}=\frac{\sqrt{x(1-x)}}{2lng(n/x)g(l/(1-x)},
\eea
and we see that the solution respects the antisymmetry of the matrix $M$.
We note that the matrix elements coupling left and right moving spin
waves factorize, unlike those coupling left to left and right to right.
\subsection{Operator Insertions}
Superstring vertices typically require insertions at the join/break point.
In terms of the string bit model, these points could be
$k=1$, $k=L$, $k=L+1$, or $k=M=L+K$. In the first two cases we can write
\bea
S_L=\frac{1}{\sqrt{L}}\sum_{n=0}^{L-1}B_n^{(1)},
\qquad S_1=\frac{1}{\sqrt{L}}\sum_{n=0}^{L-1}B_n^{(1)}e^{2i\pi n/L},
\eea
and in the last two cases
\bea
S_{L+1}=\frac{1}{\sqrt{K}}\sum_{n=0}^{K-1}B_n^{(2)}e^{2\pi in/K},\qquad S_M=\frac{1}{\sqrt{K}}\sum_{n=0}^{K-1}B_n^{(1)},
\eea
with similar expressions for ${\tilde S}$. The presence of the factor
$L^{-1/2}$ or $K^{-1/2}$ in all these expressions means that in the continuum
limit the sums over $n$ must diverge like $L^{1/2}$ or $K^{1/2}$ if
a finite contribution is to occur. We use
\bea
B_n&=&F_n\cos\frac{\pi n}{2M}+{\bar F}_n\sin\frac{\pi n}{2M}\\
i{\tilde B}_n&=&F_n\sin\frac{\pi n}{2M}-{\bar F}_n\cos\frac{\pi n}{2M},
\eea
and the divergence must come from the action of the lowering operators
on $\ket{G}$. So the possibilities are
\bea
S_L\ket{G}&\sim&\frac{1}{\sqrt{L}}\sum_{n=1}^{L-1}\cos\frac{\pi n}{2L}M_{n1,
l}f^{\dagger}_l\ket{G}\\
i{\tilde S}_L\ket{G}&\sim&\frac{1}{\sqrt{L}}\sum_{n=1}^{L-1}\sin\frac{\pi n}{2L}M_{n1,l}f^{\dagger}_l=\frac{1}{\sqrt{L}}\sum_{n=1}^L\cos\frac{\pi n^\prime}{2L}
M_{n1^\prime,l}f^{\dagger}_l\ket{G}.
\eea
In the case of $S_L$ the trig function suppresses the modes near
$n=L$, whereas for ${\tilde S}_L$ the modes near $n=0$ (or $n^\prime$ near
L) are suppressed.
As $L\to\infty$ the sum over $n$ diverges as $L^{1/2}$ so these insertions
are finite and non-zero in the continuum limit. This divergence can be
seen by considering modes in the range $1\ll n,n^\prime\ll L$.
\bea
M_{n1,l1}&\sim&\frac{1}{2}\frac{1}{\sqrt{2\pi n}}\frac{\sqrt{x}}{lg(l/x)},\qquad
M_{n2,l1}\sim-\frac{1}{2}\frac{1}{\sqrt{2\pi n}}\frac{\sqrt{x}}{lg(l/x)},\qquad
M_{n1,0}\sim\frac{1}{\sqrt{4\pi n}}\nonumber\\
M_{n1,l2}&\sim&-\frac{1}{2}\frac{1}{\sqrt{2\pi n}}\frac{\sqrt{(1-x)}}{lg(l/(1-x)},
\qquad
M_{n2,l2}\sim\frac{1}{2}\frac{1}{\sqrt{2\pi n}}\frac{\sqrt{1-x}}{lg(l/(1-x)}
\nonumber\\
M_{n2,0}&\sim&-\frac{1}{\sqrt{4\pi n}}.
\eea
The elements $M_{n^\prime,l^\prime}$ behave as the negatives of these. 
The mixed elements behave as
\bea
M_{n1,l1^\prime}&=&\frac{1}{2}\frac{1}{\sqrt{2\pi n}}\frac{\sqrt{x}}{lg(l/x)},
\qquad 
M_{n2,l1^\prime}=-\frac{1}{2}\frac{1}{\sqrt{2\pi n}}\frac{\sqrt{x}}{lg(l/x)}\\
M_{n2,l2^\prime}&=&\frac{1}{2}\frac{1}{\sqrt{2\pi n}}\frac{\sqrt{1-x}}{lg(l/(1-x)},\qquad
M_{n1,l2^\prime}=-\frac{1}{2}\frac{1}{\sqrt{2\pi n}}\frac{\sqrt{1-x}}{lg(l/(1-x)},
\eea
and from (\ref{sixth})
\bea
M_{n1^\prime,l1}&=&-\frac{1}{2}\frac{1}{\sqrt{2\pi n}}\frac{\sqrt{x}}{lg(l/x)},
\qquad 
M_{n2^\prime,l1}=\frac{1}{2}\frac{1}{\sqrt{2\pi n}}\frac{\sqrt{x}}{lg(l/x)}\\
M_{n2^\prime,l2}&=&-\frac{1}{2}\frac{1}{\sqrt{2\pi n}}
\frac{\sqrt{1-x}}{lg(l/(1-x)},\qquad
M_{n1^\prime,l2}=\frac{1}{2}\frac{1}{\sqrt{2\pi n}}
\frac{\sqrt{1-x}}{lg(l/(1-x)}.
\eea
For the insertion on string 1 at $k=L$ od $k=1$  we need
\bea
M_{n1,l}f_l^\dagger&=&\frac{1}{\sqrt{8\pi n}}\left[\sqrt{2}f_0^\dagger
+\sum_{l=1}^\infty\frac{\sqrt{x}}{lg(l/x)}(f^\dagger_{l1}+f^\dagger_{l1^\prime})
-\sum_{l=1}^\infty\frac{\sqrt{1-x}}{lg(l/(1-x))}
(f^\dagger_{l2}+f^\dagger_{l2^\prime})
\right]\nonumber\\
&\equiv&\frac{1}{\sqrt{8\pi n}}S,
\eea
and
\bea
M_{n1^\prime,l}f_l^\dagger&=&-\frac{1}{\sqrt{8\pi n}}\left[\sqrt{2}f_0^\dagger
+\sum_{l=1}^\infty\frac{\sqrt{x}}{lg(l/x)}(f^\dagger_{l1}+f^\dagger_{l1^\prime})
-\sum_{l=1}^\infty\frac{\sqrt{1-x}}{lg(l/(1-x))}
(f^\dagger_{l2}+f^\dagger_{l2^\prime})
\right]\nonumber\\
&=&-\frac{1}{\sqrt{8\pi n}}S.
\eea
The insertion of $i{\tilde S}_L$ gives the negative of the insertion of
$S_L$ because the roles of $n1$ and $n1^\prime$ are switched.
Inspection shows that all possible insertions, $S_L$, $i{\tilde S}_L$,
 $S_{L+1}$, $i{\tilde S}_{L+1}$, $S_1$, $i{\tilde S}_1$,
 $S_{M}$, $i{\tilde S}_{M}$ involve the same operator $S$ in the limit
$M,L,M-l\to \infty$, and in fact yield the same factor up to a sign. 
In this limit, the surviving part of the sum over $n$
involves
\bea
\frac{1}{\sqrt{8\pi L}}\sum_{n=n_0}^{L/2}\frac{1}{\sqrt{n}}\left(\cos\frac{n\pi}{2L}-sin\frac{n\pi}{2L}\right)
&\sim&\frac{1}{\sqrt{8\pi}} \int_{n_0/L}^1 \frac{dx}{\sqrt{x}}
\left(\cos\frac{x\pi}{2}-\sin\frac{x\pi}{2}\right)\nonumber\\
&\to&\frac{1}{\sqrt{8\pi}} \int_{0}^1 \frac{dx}{\sqrt{x}}
\left(\cos\frac{x\pi}{2}-\sin\frac{x\pi}{2}\right).
\eea
When the insertion is placed on the other string, $M-L=K$ replaces $L$,
but as long as both $L$ and $K$ are large the same result ensues. 

If we consider inserting more than one factor of $S$ at the join/break point,
we find no new operator structure in the continuum limit unless $S$ 
belongs to a different worldsheet field than the first one. 
For instance $S_k^2=1$ identically, 
If we apply say $S_kS_j$ to the overlap, with $k,j$ on the same
small string, we pass this operator through the exponential factors. If $k,j$
are at the join/break point, the commutator terms approach one or two factors
of $S$ in the continuum limit. If only one factor is produced, it will multiply
a creation term which vanishes in the continuum limit. If two factors
are produced $S^2=0$ because $S$ is an anticommuting variable, thus the
only surviving contribution in the continuum limit is the result of
applying $S_kS_j$ to the ground state of the string. The terms involving
creation operators vanish in the continuum limit, so only a $c$-number
arising from normal ordering $S_kS_j$ can survive.



\begin{thebibliography}{1}
\bibitem{goddardgrt}
  P.~Goddard, C.~Rebbi, C.~B.~Thorn,
  Nuovo Cim.\  {\bf A12 } (1972)  425-441.
P.~Goddard, J.~Goldstone, C.~Rebbi and C.~B.~Thorn,
 Nucl.\ Phys.\  B {\bf 56} (1973) 109.
\bibitem{gilest}
  R.~Giles and C.~B.~Thorn,
  Phys.\ Rev.\  D {\bf 16} (1977) 366.
\bibitem{thornsakh}
  C.~B.~Thorn,
  In *Moscow 1991, Proceedings, Sakharov memorial lectures in physics, vol. 1* 447-453, and [arXiv: hep-th/9405069].
\bibitem{bergmantsubit}
  O.~Bergman and C.~B.~Thorn,
  Phys.\ Rev.\ D {\bf 52} (1995) 5980
  [hep-th/9506125].
\bibitem{thooftlargen}
G. 't Hooft, {\sl Nucl. Phys.} {\bf B72} (1974) 461.
 \bibitem{thoofthologram}
G. 't Hooft,
{\sl Nucl. Phys.} {\bf B342} (1990) 471;
``On the Quantization of Space and Time,'' {\it Proc. of the
4th Seminar on Quantum Gravity}, 25--29 May 1987, Moscow, USSR,
ed. M. A. Markov, (World Scientific Press, 1988);
  ``Dimensional reduction in quantum gravity,''
  gr-qc/9310026.
\bibitem{sunthorn}
  S.~Sun and C.~B.~Thorn,
  Phys.\ Rev.\ D {\bf 89} (2014) 10,  105002
  [arXiv:1402.7362 [hep-th]].
\bibitem{thornspace}
  C.~B.~Thorn,
  JHEP {\bf 1411} (2014) 110
  [arXiv:1407.8144 [hep-th]].
\bibitem{greenschwarz}
  M.~B.~Green and J.~H.~Schwarz,
  Nucl.\ Phys.\ B {\bf 181} (1981) 502;
  K.~Bardakci and M.~B.~Halpern,
  Phys.\ Rev.\ D {\bf 3} (1971) 2493.
\bibitem{mandelstamlc}  S.~Mandelstam,
  Nucl.\ Phys.\  B {\bf 64} (1973) 205.
 Nucl.\ Phys.\  B {\bf 69} (1974) 77.
\bibitem{gso}
  F.~Gliozzi, J.~Scherk and D.~I.~Olive,
  Nucl.\ Phys.\ B {\bf 122} (1977) 253; 
\bibitem{rns}
P.~Ramond,
  Phys.\ Rev.\  D {\bf 3} (1971) 2415;  A.~Neveu and J.~H.~Schwarz,  
Nucl.\ Phys.\  B {\bf 31} (1971) 86; A.~Neveu, J.~H.~Schwarz and C.~B.~Thorn,
  Phys.\ Lett.\  B {\bf 35} (1971) 529.
C.~B.~Thorn,
  Phys.\ Rev.\  D {\bf 4} (1971) 1112;
A.~Neveu and J.~H.~Schwarz,  
 Phys.\ Rev.\  D {\bf 4} (1971) 1109.
\bibitem{thornfock}
  C.~B.~Thorn,
  Phys.\ Rev.\ D {\bf 20} (1979) 1435.
\bibitem{thornwsdet}
  C.~B.~Thorn,
  Phys.\ Rev.\ D {\bf 86} (2012) 066010
  doi:10.1103/PhysRevD.86.066010
  [arXiv:1205.5815 [hep-th]].
\bibitem{goldstone} J. Goldstone, private communication to S. Mandelstam,
1973.
\bibitem{greenschwarzbrink}
  M.~B.~Green, J.~H.~Schwarz and L.~Brink,
  Nucl.\ Phys.\ B {\bf 219} (1983) 437.
\end{thebibliography}
\end{document}